\documentstyle[preprint,aps,prl]{revtex}
\tighten
\draft
\begin{document}
\preprint{December 13, 1995}
\title{How do the properties of a glass depend on the cooling rate? A
computer simulation study of a Lennard-Jones system}
\author{Katharina Vollmayr\cite{kvollmayr}, Walter Kob\cite{wkob}
and Kurt Binder}
\address{Institut f\"ur Physik, Johannes Gutenberg-Universit\"at,
Staudinger Weg 7, D-55099 Mainz, Germany}
\maketitle

\begin{abstract}
Using molecular dynamics computer simulations we investigate how the
glass transition and the properties of the resulting glass depend on
the cooling rate with which the sample has been quenched. This is done
by studying a two component Lennard-Jones system which is coupled to a
heat bath whose temperature is decreased from a high temperature, where
the system is a liquid, to zero temperature, where the system is a
glass. The temperature $T_b$ of the heat bath is decreased linearly in
time, i.e. $T_b=T_0-\gamma t$, where $\gamma$ is the cooling rate.  In
accordance with simple theoretical arguments and with experimental
observations we find that the glass transition, as observed in the
specific heat and the thermal expansion coefficient, becomes sharper
when $\gamma$ is decreased.  A decrease of the cooling rate also leads
to a decrease of the glass transition temperature $T_g$ and we show
that the dependence of $T_g$ on $\gamma$ can be rationalized by
assuming that the temperature dependence of the relaxation times of the
system is given by either a Vogel-Fulcher law or a power-law.  By
investigating the structural properties of the glass, such as the
radial distribution functions, the coordination numbers and the angles
between three neighbor-sharing particles, we show how the local order
of the glass increases with decreasing cooling rate. The enthalpy and
the density of the glass decrease and increase, respectively, with
decreasing $\gamma$. By investigating the $\gamma$ dependence of
clusters of nearest neighbors, we show how these observations can be
understood from a microscopic point of view. We also show that the
spectrum of the glass, as computed from the dynamical matrix, shows a
shift towards higher frequencies when $\gamma$ is decreased.  All these
effects show that there is a significant dependence of the properties
of glasses on the cooling rate with which the glass is produced.

\end{abstract}

\narrowtext
%\twocolumn

\pacs{PACS numbers: 61.43.Fs, 61.20.Ja, 02.70.Ns, 64.70.Pf}

\section{Introduction}

If a liquid is cooled rapidly enough, so that the crystallization at or
slightly below the melting point is avoided, the final state of the
material will be an amorphous solid, i.e. a glass. This so-called
glass transition, i.e. the transition from a liquid state to an
amorphous solid state, is essentially the falling out of equilibrium of
the system, because the relaxation times of the system at the glass
transition temperature exceed the time scale of the experiment.
Therefore the resulting glass is not in thermal equilibrium. Thus it
can be expected that the properties of the glass will in general depend
on the production history of the glass, as, e.g., the cooling rate with
which the sample was cooled or, if the glass transition is pressure
driven, on the rate with which pressure was applied. That such
dependencies indeed exist has been shown in various experiments in which it
was investigated how the density of the glass, the glass transition
temperature, the specific heat or the M\"ossbauer spectrum depend on
the cooling rate~\cite{bruening92,cool_exp}.

The dependence of the properties of a structural glass on its
production history have also been studied in computer simulations. Fox
and Andersen have investigated for a one-component Lennard-Jones system
the dependence of the density on the cooling rate~\cite{fox84} and
Baschnagel {\it et al} have studied on how various properties of a
polymer glass depend on the cooling rate~\cite{baschnagel93}. However,
because of their relatively complex model, the authors of this last
work were only able to vary the cooling rate by two decades.  Lai and
Lin investigated the dependence of the Wendt-Abraham parameter on the
cooling rate~\cite{lai90} and Miyagawa and Hiwatari gave evidence that
for their binary soft sphere model the structural properties of the
glass are independent of the cooling rate, that, however, dynamical
quantities depend on the cooling rate~\cite{miyagawa89}. Also Speedy
investigated this type of question and reports that the rate with which
a system of hard spheres is compressed does not affect the density of
the glass~\cite{speedy94}.

Thus although computer simulations are in principle a very useful tool
in order to investigate the production history dependence of the
properties of glasses, since they allow to study a glass in its full
microscopic details, relatively few simulations have been done to
address this question in detail. One reason for this is that the
effects to be expected are relatively small, usually on the order of a
few percent when the cooling rate is varied by two or three
decades~\cite{bruening92,cool_exp}, and that therefore the statistical
accuracy of the data has to be rather high in order to detect these
effects at all. However, as the cited papers have shown, this problem
is not insurmountable and since the potential payoff of such types of
computer simulations, namely to understand on a {\it microscopic} level
how the production history of the glass affects its macroscopic
properties, seems to justify the effort involved.

A further reason for investigating the dependence of the properties of
the glass on the cooling rate originates from those computer
simulations in which a relatively realistic type of model is used in
order to simulate {\it real} glass formers, such as SiO$_2$. In these
simulations the quality of the applied model is in most cases judged by
its ability to reproduce some experimentally determined quantities,
such as the temperature dependence of the density, radial distribution
functions or bond angle distributions functions~\cite{kob95}. Before
such a judgement can be made, however, it is necessary to see how the
properties of the glass, as determined from a computer simulation which
involves cooling rates that can be more than ten orders of magnitude
larger than the ones in real laboratory experiments, are affected by
these high cooling rates. It might well be, that a discrepancy between
the prediction of a model and the experimental data is not due to the
inadequacy of the model, but due to the too high cooling rate with
which the glass on the computer was produced. Thus it is important to
separate these two possible sources of the discrepancy, and
investigating the cooling rate dependence of the properties of the
glass is one possible way to do this.

In the present paper we investigate the dependence of the glass
transition phenomenon and the properties of the resulting glass on the
cooling rate. The system we use is a binary mixture of Lennard-Jones
particles, since this is a prototype of a simple glass
former.  The goal of this work is to show that for this system there
are significant dependencies of the details of the glass transition and
of the properties of the glass on the cooling rate. Furthermore we try
to understand these dependencies from a microscopic point of view, thus
making use of one of the advantages of computer simulations.

The rest of the paper is organized as follows. In Sec.~II we
introduce our model and give some of the details of the simulation. In
the first part of Sec.~III we present our results regarding the
cooling rate dependence of the glass transition and in the second
part of that section we give the results of the cooling rate
dependence of the resulting glass. In Sec.~IV we then summarize and
discuss these results.

\section{Model and Details of the Simulation}
\label{secII}

The system we are investigating is a binary mixture of particles,
subsequently called type A and type B, each of them having the same
mass $m$. The interaction between two particles of type $\alpha$ and
$\beta$ ($\alpha,\beta \in \{\mbox{A,B}\}$) is given by a Lennard-Jones
potential, i.e. $V_{\alpha\beta}(r)=4\epsilon_{\alpha\beta}
[(r/\sigma_{\alpha\beta})^{12}-(r/\sigma_{\alpha\beta})^6]$. The
parameters $\epsilon_{\alpha\beta}$ and $\sigma_{\alpha\beta}$ were
chosen to have the following values: $\sigma_{AA}=1.0$,
$\epsilon_{AA}=1.0$, $\sigma_{AB}=0.8$, $\epsilon_{AB}=1.5$,
$\sigma_{BB}=0.88$ and $\epsilon_{BB}=0.5$. In order to decrease the
computational burden this potential was truncated and shifted at
$r=2.5\sigma_{\alpha\beta}$.\par

The model presented has already been used to investigate the dynamics
of strongly supercooled liquids \cite{kob_lj}. From these previous
investigations it is known that this system is a good glass former, in
that it is not prone to crystallization, even if the system is cooled
very slowly. Therefore this system is a good model for investigating the
dependencies of various properties of the glass on the cooling
rate.\par

In the following all results are given in reduced units. The unit of
length is $\sigma_{AA}$, the unit of energy is $\epsilon_{AA}$ and the
unit of time is $(m \sigma_{AA}^2/48 \epsilon_{AA})^{1/2}$.  In order
to allow the investigation of the dependence of the density of the
glass on the cooling rate and the computation of the specific heat at
constant pressure, the simulations were performed at constant pressure
$p_{ext}$.  This was done by using the algorithm proposed by Andersen
\cite{andersen80}. The external pressure $p_{ext}$ was chosen to be 1.0
and we used the value of 0.05 for $M$, the mass of the piston.

In order to simulate the cooling process we proceeded as follows. The
system was equilibrated in a $NpT$-ensemble at a high temperature $T_0$
which was for most cooling processes chosen to be 2.0 (see below for a
modification of this value).  At this temperature the relaxation times
are short and thus equilibration is fast. After this equilibration
period we began with the cooling process. For this we integrated the
equations of motion with the velocity form of the Verlet algorithm,
using a time step of 0.02. Every 150 time steps the velocities of all
the particles were replaced by new velocities which were drawn from a
Boltzmann distribution corresponding to the temperature $T_b$ of the
heat bath. The temperature of this bath was decreased linearly in time,
i.e.  $T_b(t)=T_0-\gamma t$, where $\gamma$ is the cooling rate. This
cooling process was continued until the temperature of the bath was
zero, i.e. for a time $T_0/\gamma$, which means that for the smallest
cooling rate $(\gamma = 3.125\cdot 10^{-6})$ the length of the run was
$8\cdot 10^6$ time steps. Using the multi conjugate gradient
method~\cite{press92}, the coordinates of the particles as well as the
volume were subsequently allowed to relax to the nearest local minimum
of the potential energy hypersurface. It should be mentioned that an
equivalent way to relax the configurations would have been to continue
the molecular dynamics simulation at $T_b=0$ for a {\it long} time and
that the only reason for using the conjugate gradient method was
computational efficiency.

The cooling rates we investigated were:  $\gamma=0.02$, 0.01, 0.005,
0.0025, 0.001, $5.0\cdot 10^{-4}$, $2.5\cdot 10^{-4}$, $1.0 \cdot
10^{-4}$, $5.0\cdot 10^{-5}$, $2.5\cdot 10^{-5}$, $1.25\cdot 10^{-5}$,
$6.25\cdot 10^{-6}$ and $3.125\cdot 10^{-6}$. The number of A particles
was 800 and the one for the B particles was 200. In order to test
for finite size effects we also did some runs for a system twice as
large. In order to decrease the error bars of our results we averaged
the results for each value of $\gamma$ over ten independent runs, each
of which was equilibrated at $T=2.0$ for 2000 time units.  Since
at the smallest cooling rate each of these runs took about 280 hours of
cpu time on a IBM RS6000/375 it is today probably not possible to
investigate a range of cooling rates that is significantly larger than
the one investigated here.

Finally we make a remark on the choice of the starting temperature
$T_0$ for the cooling process. It can be expected that for a given
cooling rate $\gamma$ the temperature $T_g(\gamma)$ at which the
system falls out of equilibrium, and hence also the properties of the
resulting glass, is independent of the starting temperature $T_0$, as
long as $T_0$ is sufficiently above $T_g$, or, to put it differently,
that the time constant of the cooling, $[(-dT/dt)/T_0]^{-1}=1/\gamma
T_0$, is larger than the intrinsic relaxation time of the system at the
temperature $T_0$. In order to save computer time we therefore chose
for the small values of $\gamma$ a value for $T_0$ which was lower than
the corresponding value for large values of $\gamma$. In particular we
chose $T_0$ to be 2.0 for $\gamma \geq 1.0 \cdot 10^{-4}$, equal to 1.0
for $\gamma=5.0\cdot 10^{-5}$, $2.5\cdot 10^{-5}$, $1.25\cdot 10^{-5}$
and equal to 0.5 for the two smallest values of $\gamma$. In all cases
we made sure that the value of the glass transition temperature is
indeed smaller than $T_0$, which was done by investigating the
temperature dependence of the enthalpy and the density (see below for
details).

More information regarding the details of the simulation can be found in
Ref.~\cite{vollmayr_phd}.

\section{Results}

In this section we will report the results of our simulation. In the
first subsection we will discuss the cooling rate dependence of the
system {\it during} the cooling process, i.e. look at the system at
finite temperatures, and hence investigate the details of the {\it glass
transition}.  In the second subsection we will present the results on
the cooling rate dependence of the so obtained {\it glass} at $T=0$.

\subsection{Finite Temperatures}

In this subsection we will investigate the properties of the system
{\it during} the cooling process and see how these properties depend on
the cooling rate. One of the simplest quantities one can investigate is
the enthalpy $H(T_b)=E_{kin}+E_{pot}+p_{ext}V+M\dot{V}^2/2$, where $V$
is the volume of the system. This quantity has been studied in many
previous computer simulations and it was found that on lowering the
temperature a noticeable bend in $H(T_b)$ occurs at the temperature at
which the systems falls out of equilibrium. Therefore the temperature
at which this bend occurs is usually identified with the glass
transition temperature $T_g$.

In Fig.~\ref{fig1} we show the enthalpy $H$ as a function of $T_b$, the
temperature of the bath, for all cooling rates investigated. We see
(Fig.~\ref{fig1}a) that at high temperatures the curves for the
intermediate and small cooling rates (bottom curves) fall onto a master
curve. This master curve is given by the equilibrium value of the
enthalpy of the system in its (supercooled) liquid-like state, i.e.,
by the curve one would obtain for an infinitely small cooling rate. For a
given cooling rate the enthalpy of the system will be given by this
equilibrium curve provided that the temperature of the system is high
enough to allow it to equilibrate before a significant change in
temperature, due to the cooling, has happened. If this condition is
violated the system will fall out of equilibrium, and start to show a
solid-like behavior, which is reflected by the bending of the curve for
the enthalpy. Thus the system has undergone a glass transition. 
We therefore expect that the temperature $T_g$ at which this transition
happens is smaller the lower the cooling rate is. That this is indeed
the case is shown in Fig.~\ref{fig1}b in which we show the temperature
region in which the glass transition occurs on an enlarged scale. We
recognize that the smaller the cooling rate is (curves at the bottom)
the later these curves split off from the liquidus curve. Later we
will investigate this phenomenon more quantitatively.

From Fig.~\ref{fig1}a we also recognize that for the fastest cooling
rates (top curves) the system falls out of equilibrium already at the
starting temperature $T_0$. These cooling rates are so large, that it
is no longer appropriate to see the cooling as a process which
ultimately forces the system to fall out of equilibrium because its
relaxation times exceed the time scale of the simulation, but rather as
similar to a steepest descent procedure of the system with respect to
the enthalpy. Therefore we can expect that for such large cooling rates
the temperature dependence of various quantities is qualitatively
different from the one for smaller cooling rates and the fact that the
curves for the enthalpy split off from the master curve already at
$T_0$ is an example for such a different type of behavior.

The just presented behavior for the enthalpy is very similar to the one
found for the density $\rho$. The temperature dependence of $\rho$ (see
Fig.~\ref{fig2}) shows that $\rho(T_b)$, similar to $H(T_b)$, has a
bend in the vicinity of $T_b\approx 0.4$ and that the curves
corresponding to the various values of $\gamma$ split off from the
liquidus curve at a temperature that is lower the smaller $\gamma$ is.
We also note that for this system the temperature dependence of the
density is a monotonous function of temperature.  This will allow us
later to define in a simple way a glass transition temperature. It has
to be remarked that this is not a universal property, since other
systems, such as SiO$_2$, show an anomaly in the density and thus make
the extraction of a glass transition temperature from the density much
more difficult~\cite{vollmayr_sio2}. We also recognize from the inset
in Fig.~\ref{fig2}, that for this system the density varies by more
than a factor of two when the temperature is changed from $T=2.0$ to
$T=0$, which is also in strong contrast with the case of the network
glass former SiO$_2$~\cite{vollmayr_sio2}.

By differentiating the enthalpy with respect to the temperature we
obtain $c_p$, the specific heat at constant pressure. Since our data
for $H(T_b)$ was a bit too noisy to compute its derivative directly
from a difference quotient, we approximated $H(T_b)$ by a spline under
tension~\cite{reinsch67} and then determined $c_p$ by taking the
difference quotient of this spline. In Fig.~\ref{fig3} we show the so
obtained curves for $c_p$ for all cooling rates investigated. We see
(Fig.~\ref{fig3}a) that at {\it high} temperatures and very fast
cooling rates $c_p$ drops from a value around 5 to significantly
smaller values when the temperature is increased. This decrease is
again a signature that for these large cooling rates the system falls
out of equilibrium immediately after the cooling process is started.
Thus this decrease should not be seen as the generic behavior of $c_p$
for intermediate and small cooling rates. Rather we see that for such
values of $\gamma$, $c_p$ is almost independent of $T_b$ for $T_b$
larger than 0.8. Only when the temperature is decreased to
approximately 0.5 we see that the value of $c_p$ starts to drop from a
value around 5.0 to a value around 3.0, the value expected for a
classical harmonic system.  We also recognize that even at these low
temperatures the values for $c_p$ for the fast cooling rates are
considerably larger than the harmonic value 3.0. This shows that for
these fast quenches the system behaves very anharmonic even at very low
temperatures.

In order to see the dependence of $c_p$ on the cooling rate clearer we
show in Fig.~\ref{fig3}b the temperature range in which $c_p$ shows the
strong decrease on an expanded scale. We recognize that the temperature
range in which this drop occurs is identical to the temperature range
in which the bend is observed in the enthalpy and the density (see
Figs.~\ref{fig1} and~\ref{fig2}) and thus is related to the glass
transition.

From Fig.~\ref{fig3}b we recognize that the drop of the curves
becomes sharper the smaller the cooling rate is. Thus we find that the
glass transition becomes more pronounced with decreasing cooling rate,
in accordance with the simple argument put forward by Angell {\it et
al}\cite{angell81} which says that, since the relaxation times are a
strong function of temperature, e.g., of Arrhenius or Vogel-Fulcher
type, the temperature difference $T_2-T_1$ for which the relaxation
times change by, say, a factor of 100, i.e. $\tau(T_1)/\tau(T_2)$=100,
decreases with decreasing $T_1$. Since we have seen that the glass
transition temperature $T_g$ decreases with decreasing $\gamma$, it
thus follows that the transition happens in a temperature range $T_1
\leq T_g \leq T_2$ which becomes narrower the smaller the cooling rate
is.

A qualitatively similar behavior to the cooling rate dependence of
$c_p$ is found for the cooling rate dependence of the thermal expansion
coefficient $\alpha_p(T_b)$. This quantity is given by $\alpha_p(T_b) =
V^{-1} \left .  dV(T_b)/dT_b \right |_p$ and is shown in
Fig.~\ref{fig4} for all cooling rates investigated. To compute the
derivative we made again use of splines under tension. The drop in
$\alpha_p$ at {\it high} temperatures for large values of $\gamma$ should
again be viewed as the atypical behavior of $\alpha_p$ for very large
cooling rates. For intermediate and small cooling rates we see that
$\alpha_p$ shows an almost linear dependence on temperature until it
shows a rather sudden drop when the glass transition temperature is
reached.  As in the case of the specific heat this drop becomes sharper
with decreasing cooling rate.  Thus we see also for this quantity that
the glass transition becomes more pronounced the smaller the cooling
rate is. However, it is clear that even for our smallest cooling rate
the glass transition is still smeared out over a relatively large
temperature range, unlike the real experiment at much smaller cooling
rates.

From Figs.~\ref{fig1} and \ref{fig2} we see that the temperature at
which the curve for a given value of the cooling rate splits off from
the liquid-like branch decreases with decreasing $\gamma$. This means
that the glass transition temperature $T_g$ decreases with decreasing
cooling rate. In order to investigate this effect in more quantitative
terms we use the concept of the ``fictive temperature'' as introduced
long ago by Tool and Eichlin~\cite{tool31}. This concept is founded on
two observations:  The first is that at high temperatures the
temperature dependence of most bulk properties of the system is
independent of the cooling rate, i.e. the curves for the different
cooling rates fall onto a master curve, as we see it, e.g., in
Figs.~\ref{fig1} and \ref{fig2}. This master curve can thus be
approximated by a low order polynomial and therefore extrapolated to
low temperatures. The second observation is, that the low temperature
behavior of these bulk properties, is, apart from an additive constant,
independent of the cooling rate (see, e.g., Figs.~\ref{fig1} and
\ref{fig2}). The physical reason for this is that the temperature
dependence of such bulk quantities is not very sensitive on the details
of the microscopic structure of the glass, which themselves {\it are}
dependent on the cooling rate. Therefore also this low temperature
branch can be approximated by a low order polynomial and thus
extrapolated to {\it higher} temperatures. The point where the
extrapolations of the curves at high and low temperatures
intersect is the fictive temperature which in this work we will call
the glass transition temperature $T_g$. Note that this construction is
only possible if the curves actually fall onto a master curve at high
temperatures. Since this is not the case for the fastest cooling rates
investigated here, we determined $T_g$ only for the intermediate and
small cooling rates, i.e. $\gamma \leq 0.005$.

In order to determine the low temperature master curve we shifted the
curves for the different values of $\gamma$ vertically by a $\gamma$
dependent amount until the curves lay on a master curve in the
temperature range $0\leq T_b \leq 0.2$. The so obtained master curve
was fitted in this temperature range with a quadratic polynomial. (We
note that a linear function, as has been used in several other
investigations, did not give a good fit to our low temperature data.)
The master curve for the high temperature data in the range $0.6 \leq
T_b \leq 0.8$ was fitted well by a linear function in the case of the
enthalpy, whereas a quadratic polynomial was needed in the case of 
the density.

In Fig.~\ref{fig5} we present the glass transition temperatures, as
determined by the procedure described above, as a function of the
cooling rate. The error bars are estimated from the uncertainty in the
determination of the intersection points of the extrapolated curves. We
see that, as expected, $T_g$ decreases with decreasing cooling rate.
We also recognize that the values of $T_g$ as determined from the
enthalpy are not exactly the same as the ones as determined from the
density, in that, e.g., for not too large cooling rate, the former are
systematically lower than the latter. However, at the moment we are
not able to say whether this behavior has some real underlying physical
reason or whether it just reflects some systematic error in our
determination of $T_g$. It is likely that this difference becomes
smaller at very small cooling rates, where the glass transition is
much sharper than in the present case.

The dependence of $T_g$ on the cooling rate has also been determined in
experiments~\cite{bruening92}. It was found that the
functional form

\begin{equation}
T_g(\gamma)=T_0-\frac{B}{\log(A\gamma)}
\label{eq1}
\end{equation}
is able to give a satisfactory fit to the data when $\gamma$ is varied
over 3 decades. The functional form of Eq.~(\ref{eq1}) can be
rationalized by assuming that the relaxation time $\tau(T)$ of the
system shows a Vogel-Fulcher dependence on temperature, i.e.
$\tau(T)=A\exp(B/(T-T_0))$ and by assuming that the system falls out of
equilibrium at that temperature $T_g(\gamma)$ where the relaxation time
is of the order of the inverse of the cooling rate, i.e.
$\tau(T_g(\gamma))\propto \gamma^{-1}$. In Ref.~\cite{kob_lj} it was
shown that, at constant volume, the temperature dependence of the
relaxation times of the system is also fitted very well by a power-law,
a functional form proposed by the so-called mode-coupling
theory~\cite{mct}. If we use this functional form for $\tau$, i.e.,
$\tau(T)=A/(T-T_c)^{\delta}$, we expect that $T_g$ depends on $\gamma$
via

\begin{equation}
T_g(\gamma)=T_c+(A\gamma)^{1/\delta}\quad.
\label{eq1b}
\end{equation}

Also included in Fig.~\ref{fig5} are fits to the data with the
functional forms given by Eq.~(\ref{eq1}) (solid curves) and
Eq.~(\ref{eq1b}) (dashed curves). We see that both fits are able to
represent the data very well. The temperature $T_0$ from
Eq.~(\ref{eq1}), i.e., the glass transition temperature for an
infinitely slow cooling process, are 0.334 and 0.348 for the enthalpy
and the density, respectively and the corresponding values for $T_c$
from Eq.~(\ref{eq1b}) are 0.378 and 0.386, respectively.  Also in this
case it is not clear whether the difference between the values of $T_0$
(or $T_c$) for $H$ and $\rho$ has some underlying physical reason or
whether it is just a systematic error in the way we determined
$T_g(\gamma)$.

We also note that the estimated value for $T_c$ is significantly
larger than the one estimated for $T_0$. The reason for this is that
the critical temperature $T_c$ of mode-coupling theory is expected to
overestimate the glass transition temperature since the theory
neglects certain types of relaxation processes which restore
ergodicity. Thus in experiments it is found that at the temperature
$T_c$ the system shows only a {\it smeared out} singularity and that
$T_c>T_0$.

\subsection{Zero Temperatures}

After having discussed in the preceding subsection the cooling rate
dependence of various quantities at {\it finite} temperatures we now
investigate how the properties of the glass at {\it zero} temperature
depend on the cooling rate.

We have seen in Fig.~\ref{fig1} that the smaller the cooling rate is
the lower is the temperature at which the curve for the corresponding
$\gamma$ splits off from the liquidus curve. Thus we expect that the
value of the enthalpy at zero temperature will depend on $\gamma$. That
this is indeed the case is shown in Fig.~\ref{fig6}, where we plot
$H_f$, the final enthalpy after the quench (and the subsequent
relaxation of the system, see Sec.~\ref{secII}) as a function of
$\gamma$. As expected, we find that $H_f$ decreases with decreasing
cooling rate. We see that the dependence of $H_f$ on $\gamma$ becomes
stronger when we change $\gamma$ from small values to intermediate and
large values. For very large values the dependence becomes, however,
weaker again. This can be understood by noticing, that for such large
values of $\gamma$ the cooling process is very similar to a steepest
descent procedure and that therefore in the limit $\gamma \to \infty$
the dependence of the final state on the
cooling rate vanishes.

Also included in Fig.~\ref{fig6} is the result of a cooling process for
$\gamma=5.0\cdot10^{-5}$ for a system size of 2000 particles. This
point was generated by averaging over 5 different initial
configurations and the goal of this ``experiment'' was to see whether
our results are affected in some way by finite size effect. Since the
result for the system sizes $N=1000$ and $N=2000$ are the same to
within the error bars, we conclude that, for the types of questions
investigated in this work, finite size effects are not important for
the model studied here.

Since the system will try to decrease its enthalpy during the quench,
the cooling process can also be viewed as an optimization process in
which the system tries to minimize a cost function, e.g., the
enthalpy.  The dependence of the value of the cost function on the
amount of time spent to find this value has been the subject of quite a
few earlier investigations. A discussion of some of this previous work
can be found in Ref.~\cite{kob90} (see also
\cite{baschnagel93,grest_huse,new_optimization}).

In the just mentioned investigations dealing with cooling or
optimization processes it has been proposed that the dependence of the
cost function, i.e. in our case the enthalpy, on the cooling rate is
given by a power law or by a logarithmic dependence, i.e.

\begin{equation}
H_f(\gamma)=H_f^0+a_1 \gamma^{a_2}
\label{eq2}
\end{equation}

or

\begin{equation}
H_f(\gamma)=H_f^0+b_1(-\log \gamma)^{b_2} \quad ,
\label{eq3}
\end{equation}
where $H_f^0$, $a_i$ and $b_i$ are fit parameters~\cite{grest_huse}.
Therefore we have tried to fit our data for $H_f(\gamma)$ with these
functional forms and the best fits are included in Fig.~\ref{fig6} as
well. Since the functional forms can be expected to hold only for small
cooling rates, the three largest cooling rates were not included in the
fitting procedure. These cooling rates are the ones for which the
dependence of $H_f$ on $\gamma$ becomes weak again, i.e. for which the
cooling process is similar to a steepest descent. Figure~\ref{fig6}
shows that {\it both} functional forms are able to fit the data very
well. Thus within the accuracy of our data and the cooling range
investigated it is not possible to decide which functional form, if
any, is the correct one.

We also notice that for cooling rates $\gamma \leq 2.5\cdot 10^{-4}$ the
data points for $H_f$ lie, within the accuracy of our data, on a
straight line. It is clear that such a functional dependence of $H_f$
on $\gamma$ cannot be valid for arbitrarily small cooling rates, since
this would lead to arbitrarily low values of $H_f$ for sufficiently
small $\gamma$. Thus if there actually exists a regime for which there
is a linear dependence of $H_f$ on the logarithm of $\gamma$, there
also must exist at even smaller values of $\gamma$ a further regime, in
which $H_f$ shows the real asymptotic dependence on $\gamma$.  However,
this regime is currently outside of our computational possibilities.

In Fig.~\ref{fig3}a we have shown that for large cooling rates the
system behaves very anharmonic even at low temperatures. Now we find
that for these cooling rates the enthalpy of the local minimum in which
the systems ends at $T_b=0$ is relatively large. Thus we conclude that
for this system the hypersurface of the enthalpy has the property that
on average its shape is very anharmonic when the enthalpy is large and
relatively harmonic when the enthalpy is small.

Similar to the enthalpy we have found that also $\rho_f$, the density
of the system after the quench, shows a small but noticeable dependence
on the cooling rate. The results are shown in Fig.~\ref{fig7}. Note
that a change of $\gamma$ by about four orders of magnitude leads to a
change in $\rho_f$ of only 2\%, a figure which is comparable to the one
found in experiments (although at vastly slower cooling
rates)\cite{bruening92}. The smallness of the effect shows that it is
necessary to vary the cooling rate over an extensive range and to use
system sizes that are not too small in order to obtain results of
sufficient statistical accuracy.

Also in the case of $\rho_f$ we found that, for not too large values of
$\gamma$, the data can be fitted by the two functional forms given by
Eqs.~(\ref{eq2}) and (\ref{eq3}) equally well. The densities $\rho_f^0$
are 1.246 and 1.253 for the power-law and logarithmic functional forms,
respectively. Furthermore we see, that, as in the case of the enthalpy,
also for this quantity finite size effects are not important and that
the data points for intermediate and small values of $\gamma$ lie on a
straight line, thus indicating the possibility of a further type of
relaxation regime at even lower values of $\gamma$.

Since we have now seen that bulk quantities such as the enthalpy and
the density show a dependence on the cooling rate, it is interesting to
investigate how these dependencies can be understood from a
microscopic point of view. Therefore we will investigate in the
following how the microscopic structure of the glass changes when
the cooling rate is varied. In Fig.~\ref{fig8} we show the radial
distribution functions $g_{AA}(r)$ and $g_{BB}(r)$ for the AA and BB
correlations~\cite{hansen_mcdo}. The results for $g_{AB}(r)$ are
similar to the ones for $g_{AA}(r)$. The inset in Fig.~\ref{fig8}a
shows that the overall form of $g_{AA}(r)$ does not show a {\it strong}
dependence on $\gamma$. A closer investigation of the first nearest
neighbor peak showed that for the smallest cooling rate its height is
about 2\% percent higher than the one for the largest cooling
rate~\cite{vollmayr_phd}. A somewhat stronger dependence can be
observed in the second nearest neighbor peak, part of which is shown in
the main figure of Fig.~\ref{fig8}a.  We recognize that the curve for
the slowest cooling rate (solid bold line) shows more pronounced peaks
than the curve for the fastest cooling rate (dashed bold line).  Thus
we conclude that the local order of the system, as measured by
$g_{AA}(r)$ becomes more pronounced with decreasing cooling rate.

A similar behavior as the one observed for $g_{AA}(r)$ is found for
$g_{BB}(r)$ (Fig.~\ref{fig8}b). However, in this case the height of the
{\it first} nearest neighbor peak shows a relatively strong dependence
on the cooling rate. We see that in this case the height of this peak
{\it decreases} with decreasing cooling rate.  This behavior can be
understood by remembering that $\epsilon_{BB}$, the interaction energy
of the Lennard-Jones potential between two B particles, is only $0.5$,
as compared to 1.5 for $\epsilon_{AB}$.  Therefore the system will try
to avoid to have two B particles as nearest neighbors, and to have A
and B particles as nearest neighbors instead, and it will manage to do
this better the more time it is given to do so, i.e. the smaller the
cooling rate is~\cite{vollmayr95}.  However, from the figure we also
recognize that the height of the first nearest neighbor peak decreases
when the cooling rate is decreased from very large values to intermediate
values, but that for even smaller cooling rates the $\gamma$-dependence
of the height is much weaker, if it exists at all. This indicates that the
above mentioned driving force for decreasing the peak is no longer
effective for intermediate and small cooling rates and that therefore a
different driving force must exist in order to explain the decrease of
the enthalpy or the density.

Having the information on the various radial distribution functions we
can now investigate the dependence of the different types of
coordination numbers of the particles on the cooling rate.  We define
the coordination number $z$ of a particle of type $\alpha$ to be the
number of particles of type $\beta$ that are closer to the first
particle than $r_{\min}^{\alpha\beta}$, the location of the first
minimum in $g_{\alpha\beta}(r)$. We have found that the value of
$r_{min}^{\alpha\beta}$ depends only weakly on the cooling rate~\cite{vollmayr_phd} and
thus we have chosen the following, $\gamma$-independent, values for
$r_{min}^{\alpha\beta}$:  $r_{min}^{AA}=1.4$, $r_{min}^{AB}=1.2$ and
$r_{min}^{BB}=1.07$.

In Fig.~\ref{fig9} we show the $\gamma$ dependence of
$P_{\alpha\beta}$, the probability that a particle of type $\alpha$ has
$z$ nearest neighbors of type $\beta$ for various values of $z$.  We
see that the curves for the AA pairs, Fig.~\ref{fig9}a, for $z=12$ and
$z=13$ increase with decreasing $\gamma$, that the curve for $z=10$
shows a decreasing trend and that the curve for $z=11$ is essentially
independent of $\gamma$. We recognize that the changes that take place
in the distribution of the coordination number are relatively small,
i.e. less than 10\%, when the two particles are of type A. This is not
the case for the BA and BB pairs as is shown by Figs.~\ref{fig9}b and
c. Here we see that the change in the coordination number can be as
much as 50\% ($z=7$ and $z=9$ in Fig.~\ref{fig9}b). Furthermore we find
that the coordination numbers for BB pairs for $z=0$ and $z=1$ increase
and decrease, respectively, with decreasing cooling rate
(Fig.~\ref{fig9}c). This observation is in accordance with the comments
we made in the context of Fig.~\ref{fig8}b, where we found that the
height of the first nearest neighbor peak in the BB correlation
function is decreasing with decreasing $\gamma$. We also see in
Fig.~\ref{fig9}c that for values of $\gamma$ less than $10^{-4}$, the
value of $P_{BB}$ for $z=0$ is essentially independent of $\gamma$,
which is in accordance with the comments made before. We also mention
that from Fig.~\ref{fig9}a one should {\it not} conclude that, since
the curve $P_{AA}$ for $z=12$ shows a increasing trend, for very small
cooling rates the {\it total} number of nearest neighbor particles of a
A particle is 12. What is shown in the figure is just the number of
nearest A particles, thus the B neighbors are not taken into account.
As we will show below the most frequent values of the total
coordination number for the A particles is 13 and 14. This is different
from the results one would expect for a one-component system, where at
low temperatures particles tend to pack locally in an icosahedral
structure, thus leading to a coordination number of 12. Since in our
case we have a mixture of large and small particles it can thus be
expected that the value of the most frequent coordination number is
larger than 12 and below we will show that this is indeed the
case~\cite{clusters}.

The fact that the nearest neighbor particles of an A particles do not
form an icosahedron is also supported by the form of the bond-bond
angle distribution function. Here we define a bond as the line segment
connecting two neighboring particles and a bond-bond angle as the angle
between two bonds which are connected to a common particle.  In
Fig.~\ref{fig10} we show this distribution function for the case of
three A particles.  Note that if the local structure of the particles
would be a perfect icosahedron we would have only contributions at
angles $63.4^{\circ}$ and $116.6^{\circ}$.  We see (inset in
Fig.~\ref{fig10}), that the distribution function shows indeed maxima
at the mentioned angles, but that the peaks are very broad, thus
showing that the local structure is not an icosahedron. The overall
form of the distribution function shows only a small dependence on the
cooling rate, the most pronounced effect seems to be that the height of
the peak near $60^{\circ}$ increases and shows the tendency to shift
its position to larger values of the angle with decreasing cooling rate
(main figure). The smallness of these effects is in stark contrast with
our findings for the network glass former SiO$_{2}$ for which we found
a strong dependence of the distribution function for the angles on the
cooling rate\cite{vollmayr_sio2}.

Since we have seen now that the arrangement of the particles making up
the nearest neighbor shell of the A particles is not similar to an
icosahedron, we now investigate what the nature of this arrangement
really is. Furthermore we will also try to understand, what the
underlying reason for the cooling rate dependence of $H_f$ and $\rho_f$
really is.  In order to do this we introduce the notion of a
``cluster''. We define a cluster of particles as the collection of
particles given by a central particle and the particles of its nearest
neighbor shell (computed by using the distances
$r_{min}^{\alpha\beta}$). We say that a cluster is of type
$\alpha_{\mu,\nu}$ if the central particle is of type $\alpha\in
\{\mbox{A,B}\}$ and if it has $\mu$ nearest neighbors, $\nu$ of which
are of type B. Note that this definition of a cluster can be viewed as
a first step of a whole hierarchy of definitions of clusters in which,
e.g., one defines clusters as the collection of particles that are
within the first or first two nearest neighbor shell(s), etc. and
specifying how many particles of each type are in any given shell.
However, in order to avoid to complicate the analysis of the data too
much we restrict ourselves to this simplest kind of cluster
definition.

We define the energy of a cluster to be given by the sum of all
pairwise interactions between any two members of the cluster and will
denote this energy by $E_c$. The dependence of $E_c$ on the cooling
rate is shown in Fig.~\ref{fig11} for various types of clusters. In
order not to crowd the figure too much, we show only a representative
selection of the types of clusters we found. (More curves can be found
in Fig.~2 of Ref.~\cite{vollmayr_epl}). The dependence of $E_c$ on
$\gamma$ for the types of clusters not shown is very similar to the one
of the types of clusters shown. The following observations can be
made:  1) For all types of clusters $E_c$ decreases with decreasing
cooling rate.  A variation of 4 decades in $\gamma$ gives rise to a
decrease of about 1\%. 2) For a given cooling rate the difference of
$E_c$ between the different types of clusters is usually much larger
than 1\%. 3) For a given number of B particles in a cluster, $E_c$
decreases with increasing number of $A$ particles in the cluster (as
exemplified in the figure by the clusters A$_{\mu,2}$ (filled
symbols)). 4) For a given total number of particles in the cluster,
$E_c$ decreases with increasing number of B particles (as exemplified
in the figure by the clusters A$_{14,\nu}$ (open symbols)). Although we
have shown here only the dependence of $E_c$ on $\gamma$ for clusters
of type A$_{\mu,\nu}$ similar results hold true also for clusters of the
type B$_{\mu,\nu}$~\cite{vollmayr_phd}.

Since for each type of cluster a variation of $\gamma$ by four decades
gives rise to a change in $E_c$ on the order of one percent and since the
change of the enthalpy, which is related to the sum of the energy of
the clusters, is also on the order of 1-2\%, one might conclude that
the change of the enthalpy can be rationalized solely by the change of
the energy of the clusters. Before one can draw this conclusion one
has, however, to see whether the distribution of the frequency of the
different types of clusters is not also changing with the cooling
rate.  In Fig.~\ref{fig12} we show the probability
$P_{{\alpha}_{\mu,\nu}}$ that a cluster is of type $\alpha_{\mu,\nu}$
as a function of $\gamma$. In order not to crowd the two figures too
much, we show only those curves for which this probability is not too
small.  We recognize that there are clusters for which this probability
is essentially independent of $\gamma$, such as, e.g., cluster
A$_{13,2}$ in Fig.~\ref{fig12}a, and that there are quite a few
clusters for which this probability changes by more than 10\% (all the
clusters that are marked by symbols).  Since we have seen from
Fig.~\ref{fig11} that the difference between the energy of different
clusters is usually much larger than 1-2\%, the amount of energy a
particular cluster changes when $\gamma$ is changed by four decades, we
thus find that the change in the distribution of the frequency of the
clusters might be the much more important mechanism for lowering the
enthalpy than the lowering of the energy of an individual cluster.

Before we address this question further we briefly comment on a
different observation. From Fig.~\ref{fig12} we recognize, that the
observation we made in in Fig.~\ref{fig9}a, namely that the curve for
the AA coordination number $z=11$ is independent of the cooling rate,
does not imply that the composition of the clusters containing 11 A
atoms is independent of $\gamma$. Rather we see that the curves for
cluster types A$_{14,3}$ and A$_{12,1}$, both of them having 11 A
particles, show a quite strong dependence on $\gamma$. However, this
dependence is such that they essentially cancel each other, i.e., that
the $\gamma$ dependence of the abundance of the clusters containing 11
A particles is almost zero.

In order to understand the relative importance of the two above
mentioned mechanisms to lower the enthalpy for the system with
decreasing cooling rate, i.e. the lowering of the energy of each
individual cluster and the changing of the distribution function
$P_{\alpha_{\mu,\nu}}$, it is useful to look at $P(E_c)$, the
probability that a cluster has an energy $E_c$. In Fig.~\ref{fig13} we
show this probability for all cooling rates investigated.  We see that
this distribution shows various peaks, each of which can be related to
certain types of clusters (see labels). It should be noted that the
{\it position} of a peak is related to the average energy $E_c$ of the
corresponding cluster and that the {\it height} of a peak is related to
the frequency with which it occurs. We recognize from the figure that
the main difference between the distribution for the slowest cooling
rate (solid bold curve) and the fastest cooling rate (dashed bold
curve) is that the height of the various peaks is significantly
different. The frequency of cluster types, i.e. the value of $P(E_c)$,
with high energies is decreasing and the one of cluster types with low
energies is increasing with decreasing $\gamma$. The location of the
peaks, however, shows only a very small change in the direction of
lower energies, in accordance with point 1) in our discussion of
Fig.~\ref{fig11}. Thus we come to the conclusion that {\it main} reason
for the observed decrease of the enthalpy of the system with decreasing
cooling rate is not that the energy of the individual clusters is
decreasing, but that the distribution of the frequency with which these
clusters occur shows a relatively strong dependence on the cooling
rate.

Similar to our definition of the energy of a cluster it is also
possible to introduce the density of a cluster. Therefore it is
possible to investigate how the distribution of the density of the
clusters depend on the cooling rate and we found that, as it is the
case for the enthalpy, it is the cooling rate dependence of the
distribution of the clusters, rather than the density of the clusters
themselves, that is the main reason for the observed cooling rate
dependence of the total density of the system~\cite{vollmayr_phd}.

To close the subsection containing the results of our analysis of the
cooling rate dependence of the glass at zero temperature, we present
the results of our investigation about the spectrum of the glass. In
order to gain some insight into the nature of the so-called boson peak
(see, e.g., Ref.~\cite{physica}), the low-temperature spectrum of
glasses has in recent years been the focus of several investigations in
which it was attempted to address this question by means of computer
simulations~\cite{laird_schober,olig_bermej,bembenek95}.  Laird and
Schober showed that in a glass in which the particles interact with a
soft sphere potential there exist localized low-frequency modes which
might be related to the boson peak\cite{laird_schober}, a result which
was also found for a model glass for Selenium~\cite{olig_bermej}, and
recently Bembenek and Laird have also investigated the connection
between the shape of the spectrum and the glass
transition~\cite{bembenek95}. In view of this work it is therefore
interesting to see how the spectrum is affected when the cooling
rate with which the glass was produced is changed.

We determined the spectrum of a given relaxed configuration at $T=0$ by
computing the eigenvalues of the dynamical matrix, i.e of $\partial^{2}
V(\{ {\bf{r}}_i\})/\partial r_{j,\alpha} \partial r_{k,\beta}$, where
$j$ and $k$ are particle indices and $\alpha$ and $\beta$ are the
cartesian components $x,y,z$. From each configuration we thus obtained
$3N$ eigenvalues $\lambda$, where $N$ is the total number of particles,
i.e., 1000. From these eigenvalues, all of which are positive, since
the configuration is locally stable, we computed the frequencies
$\nu=\sqrt{\lambda/m}/2\pi$. In Fig.~\ref{fig14} we show the so
obtained spectrum $Z(\nu)$ for all cooling rates investigated. Not
included in this figure are the three trivial eigenvalues of zero which
correspond to a mere translation of the system.  A comparison of this
figure with the spectra shown in Ref.~\cite{olig_bermej} for
Se shows that the latter spectra are more structured, i.e. show more
peaks than the one for the Lennard-Jones mixture. This difference is
probably due to the fact that the potential of the models investigated
by the authors of Ref.~\cite{olig_bermej} try to mimic directional
bonds, whereas the Lennard-Jones model investigated here is more
similar to a hard sphere model with the corresponding packing
structure. This argument is given further support by our observation
that also the network forming glass SiO$_2$ shows a spectrum which is
much more structured than the one for the Lennard-Jones system
presented here~\cite{vollmayr_sio2}.

We recognize from Fig.~\ref{fig14} that the spectrum shows only a weak
dependence on the cooling rate. That there actually is, however, such a
dependence is shown in Fig.~\ref{fig15}a, where we show an enlargement
of the low frequency part of the spectra. We clearly see that with
decreasing cooling rate the spectrum moves to higher frequencies. A
similar behavior is observed for the high frequency wing of the
spectrum. Thus we come to the conclusion that with decreasing cooling
rate the average local environment of the particles transforms in such
a way that the local potential of the particles becomes stiffer. Such
an effect is not implausible since we have seen that the density of
the system increases with decreasing cooling rate and that therefore
the particles are closer packed.

In order to estimate how large of an effect this shifting of the
spectrum to higher frequencies is we proceeded as follows. From
Fig.~\ref{fig15}a we recognize that on the low frequency wing of the
spectrum the main effect of the cooling rate seems to be that the
spectrum is shifted to the right, i.e., without a change of the form of
the curves. We therefore made use of this observation and attempted to
shift the curves that have a value of $\gamma$ larger than $3.1\cdot
10^{-6}$, the smallest cooling rate, {\it horizontally}, i.e., in
frequency, by an amount $\Delta\nu(\gamma)$ such that they collapsed as
well as possible with the curve for $\gamma=3.1\cdot 10^{-6}$ in the
range where the values of $Z(\nu)$ is between 0.2 and 1.0.  The quality
of the collapse between two curves was determined with the Spearman
rank order correlation coefficient~\cite{press92} and the shift
$\Delta\nu(\gamma)$ was determined by maximizing this coefficient. In
Fig.~\ref{fig15}a we show the so obtained master curve (right set of
curves).  (In order to make the figure clearer we shifted this master
curve by 0.02 to higher frequencies.) We thus recognize that it is
indeed possible to collapse the curves for the different cooling rates
onto one master curve, which shows that our procedure to estimate the
frequency shift of the spectra as a function of the cooling rate is
reasonable. In Fig.~\ref{fig15}b we show the amount of shifting
$\Delta\nu(\gamma)$ as a function of $\gamma$. We see that there is a
clear trend that $\Delta\nu$ decreases with decreasing $\gamma$, i.e.,
that the low frequency part of the spectra moves to higher frequencies.
It is interesting to note that the curve $\Delta\nu(\gamma)$ is
approximated quite well by a straight line, i.e., that there is no
indication that the spectrum converges towards an asymptotic
distribution with decreasing $\gamma$. This means that with respect to
this quantity we are not yet seeing the asymptotic behavior expected
for very small values of $\gamma$, thus giving us evidence that there
might be, at even smaller values of $\gamma$, a crossover from the
$\gamma$ dependence of $\Delta\nu$ we are observing here, to a
different one. Note that this possibility is also in accordance with
our finding in the context of the enthalpy of the system at $T=0$,
where we also had some evidence that there might be a crossover at even
smaller cooling rates than the ones considered here.

Although the effect observed here is relatively small, and therefore
will probably not affect the conclusions drawn in the work of
references~\cite{laird_schober,olig_bermej,bembenek95}, we mention that
a similar investigation for the strong glass former SiO$_2$ has shown
that there are systems where the spectrum is strongly affected by the
cooling rate~\cite{vollmayr_sio2}. Thus such effects should probably be
taken into account in the future.

Apart from this cooling rate dependence of the spectrum we recognize
from Fig.~\ref{fig14} that $Z(\nu)$ also has a system size dependence,
since the spectrum shows a gap at low frequencies. One knows that
acoustic phonons with long wavelength do exist in a glass, as they
exist in a crystalline solid. Phase space arguments imply that $Z(\nu)
\propto \nu^2$ for $\nu \to 0$. This is not found in the simulation,
since the finite linear dimension $L$ of the box implies that the
largest phonon wavelength is given by $L$. Hence the corresponding
frequency is the lowest frequency occurring in the spectrum. Thus this
finite size effect should be taken into account if one simulates very
small systems.

\section{Summary and Conclusions}

We have performed a molecular dynamics computer simulation of a binary
Lennard-Jones system in order to investigate the dependence of the
properties of a glass on its production history. In order to do this we
cooled the system, with a cooling rate $\gamma$, from its high
temperature liquid phase to a glass at zero temperature, and
investigated how the glass transition as well as the so produced glass
depend on this cooling rate.

In qualitative accordance with experiments~\cite{bruening92,cool_exp}
we find that the glass transition becomes sharper with decreasing
cooling rate, which is also in accordance with simple arguments put
forward by Angell {\it et al}~\cite{angell81}.  In particular we see,
that in the glass transformation range, the specific heat as well as
the thermal expansion coefficient show a more rapid change the smaller
the cooling rate is.  Also the glass transition temperature $T_g$ shows
a cooling rate dependence and we demonstrate that this dependence can
be understood by assuming that the temperature dependence of the
relaxation times of the system is given by either a Vogel-Fulcher law
or a power-law.

Our analysis of the glass shows that its density, its enthalpy as well
as its structural properties, such as the radial distribution function,
the coordination numbers and bond-bond angles, show a relatively small,
but nevertheless noticeable (on the order of a few percent) dependence
on the cooling rate with which the glass was produced. By investigating
the cooling rate dependence of nearest neighbor clusters we give
evidence that the dependence of quantities like the enthalpy or the
density is mainly caused by the cooling rate dependence of the
distribution of the frequency with which the various types of clusters
occur.

Finally we have also investigated how the spectrum $Z(\nu)$ of the
glass depends on the cooling rate. We find that $Z(\nu)$ shows only a
weak dependence on $\gamma$ and that the main effect seems to be that
the whole spectrum is shifted towards higher frequencies, a behavior
which is quite plausible. It is interesting to note that the amount
$\Delta\nu(\gamma)$ by which the spectrum is shifted, does not show any
tendency to saturate within the range of cooling rates that we are
able to access. Therefore it seems that the range of cooling rates
falls at least into three regimes. The first one includes the very fast
cooling rates. In this regime the cooling process is essentially a
steepest descent procedure and the dependence of the properties of the
glass on the cooling rate is relatively weak. In the second regime
some quantities, such as the glass transition temperature or the
density, show a {\it relatively} strong dependence on the cooling rate,
but this dependence seems to become weaker with decreasing cooling
rate and thus an extrapolation to experimental values of the cooling
rate seems to be feasible. However, there seem to be quantities, such
as the shift $\Delta\nu(\gamma)$ of the spectrum or, to a lesser extent,
the enthalpy of the system, which show in this second regime a
cooling rate dependence which is not possible for arbitrarily small
values of $\gamma$. Therefore we conclude that there must be a third
regime of cooling rates in which {\it all} quantities actually show a
cooling rate dependence which is the asymptotic one.

To conclude we can say that this work has shown that also in computer
simulations the phenomenon of the glass transition, as well as the
properties of the glass, show a clear dependence on the cooling rate
and that these dependencies can be understood also from a microscopic
point of view.  For the model investigated here these effects are
relatively small but we know that they might be substantial in other
types of systems~\cite{vollmayr_sio2}. Therefore it is advisable that
in future simulations which investigate the glass transition, this
aspect is not left out since, e.g., the outcome of investigations in
which it is tested whether a specific model is able to reproduce
experimental properties of glasses, might be severely affected by the
history on how the glass was produced on the computer.

Acknowledgements: We thank C. A. Angell for valuable discussions. K.
V. thanks Schott-Glaswerke, Mainz and the DFG, through SFB 262, for
financial support. Part of this work was done on the computer
facilities of the Regionales Rechenzentrum Kaisers\-lautern.

\newpage
\begin{figure}
\caption{Enthalpy $H$ of the system versus $T_b$, the temperature of
the heat bath, for all cooling rates investigated.  a) Full range of
temperature. b) Enlargement of the glass transition region. The solid
and dashed bold curves are the smallest and largest cooling rates,
respectively.
\protect\label{fig1}}
\vspace*{5mm}
\par

\caption{Density $\rho$ of the system versus $T_b$, the temperature of
the heat bath, for all cooling rates investigated. Main figure: The
glass transition region. The solid and dashed bold curves are the
smallest and largest cooling rates, respectively.  Inset: Full range of
temperature. \protect\label{fig2}}
\vspace*{5mm} 
\par

\caption{Specific heat at constant pressure versus $T_b$ for all
cooling rates investigated. The solid and dashed bold curves are the
smallest and largest cooling rates, respectively. a) Full range of
temperature. b) Enlargement of the glass transition region.
\protect\label{fig3}}
\vspace*{5mm}
\par

\caption{Thermal expansion coefficient $\alpha_p$ versus $T_b$
for all cooling rates investigated. The solid and dashed bold curves
are the smallest and largest cooling rates,
respectively.\protect\label{fig4}}
\vspace*{5mm}
\par

\caption{Glass transition temperature $T_g$, as determined from
the enthalpy (circles) and the density (diamonds), versus the cooling
rate. The solid and dashed lines are fits with the functional forms given in
Eq.~(\protect\ref{eq1}) and Eq.~(\protect\ref{eq1b}), respectively.
\protect\label{fig5}}
\vspace*{5mm}
\par

\caption{$H_f$, the value of the enthalpy after the quench, as a
function of the cooling rate for the system of 1000 particles (open
symbols) and the system of 2000 particles (filled symbol). Also
included are the results of fits with the functional forms given in
Eq.~(\protect\ref{eq2}) (dashed curve) and Eq.~(\protect\ref{eq3})
(solid curve).\protect\label{fig6}}
\vspace*{5mm}
\par

\caption{$\rho_f$, the value of the density after the quench, as a
function of the cooling rate for the system of 1000 particles (open
symbols) and the system of 2000 particles (filled symbol). Also
included are the results of fits with the functional forms given in
Eq.~(\protect\ref{eq2}) (dashed curve) and Eq.~(\protect\ref{eq3})
(solid curve).\protect\label{fig7}}
\vspace*{5mm}
\par

\caption{The radial distribution function at $T=0$ for all cooling 
rates investigated. The solid and dashed bold line correspond to the
slowest and fastest cooling rates, respectively. a) $g_{AA}(r)$. b)
$g_{BB}(r)$. \protect\label{fig8}}
\vspace*{5mm}
\par

\caption{Probability that a particle has a coordination number $z$ as a
function of the cooling rate.  a) AA pairs, b) BA pairs and c) BB
pairs. \protect\label{fig9}}
\vspace*{5mm}
\par

\caption{Distribution of the bond-bond angle between three A particles
for all cooling rates investigated. Inset:  The whole distribution.
Main figure: Enlargement of the peak around $60^{\circ}$. The solid and
dashed bold lines correspond to the fastest and slowest cooling rate,
respectively.\protect\label{fig10}}
\vspace*{5mm}
\par

\caption{$E_c$, the energy of a cluster as a function of the cooling
rate for selected cluster types. Note that the energy of the cluster of
type $A_{\mu,\nu}$ decreases with increasing $\mu$ when $\nu$ is held
fixed (filled symbols) and that $E_c$ also decreases with increasing
$\nu$ when $\mu$ is held fixed (open symbols).\protect\label{fig11}}
\vspace*{5mm}
\par

\caption{Probability to find a cluster of a given type as a function of
$\gamma$. Only the curves for the clusters showing a noticeable change
as a function of $\gamma$ (marked with symbols) or that are very
frequent are labeled. a) Clusters around A particles. b) Clusters
around B particles.\protect\label{fig12}}
\vspace*{5mm}
\par

\caption{Probability that a cluster has an energy $E_c$ for all
cooling rates investigated. The solid and dashed bold lines correspond
to the smallest and largest cooling rate, respectively. The labels of
the peaks identify the type of clusters that give rise to the
corresponding peak. \protect\label{fig13}}
\vspace*{5mm}
\par

\caption{$Z(\nu)$, the spectrum of the system at $T=0$ for all
cooling rates investigated. \protect\label{fig14}}
\vspace*{5mm}
\par

\caption{a) Low frequency part of the spectrum $Z(\nu)$ for all cooling
rates investigated (left set of curves). The solid and dashed bold
lines correspond to the smallest and largest cooling rate,
respectively. The right set of curves are the same $Z(\nu)$ but now
shifted by $\Delta\nu(\gamma)+0.02$ in order to collapse them onto a
master curve (see text for details). b) The shift $\Delta\nu(\gamma)$
as a function of the cooling rate.\protect\label{fig15}}
\vspace*{5mm}
\par

\end{figure}

\begin{references}
\bibitem[\dag]{kvollmayr}
Electronic mail: vollmayr@moses.physik.uni-mainz.de
%
\bibitem[*]{wkob}
Corresponding author.
Electronic mail: kob@moses.physik.uni-mainz.de\\
http://www.cond-mat.physik.uni-mainz.de/\~{ }kob/home\_kob.html
%
\bibitem{bruening92}
R. Br\"uning and K. Samwer, Phys. Rev. B, {\bf 46}, 11318 (1992);
%
\bibitem{cool_exp}
see, e.g.,
H. N. Ritland, J. Amer. Ceram. Soc. {\bf 37}, 370 (1954);
%
C. Y. Yang, D. E. Sayers and M. A. Paesler, Phys. Rev. B {\bf 36},
8122 (1987);
%
C. T. Limbach and U. Gonser, J. Non-Cryst. Solids {\bf 106}, 399
(1988);
%
G. P. Johari, A. Hallbrucker and E. Mayer, J. Phys. Chem. {\bf 93},
2648 (1989);
%
R. Br\"uning and M. Sutton, Phys. Rev. B {\bf 49}, 3124 (1994).
%
\bibitem{fox84}
J. R. Fox and H. C. Andersen, J. Phys. Chem. {\bf 88}, 4019 (1984).
%
\bibitem{baschnagel93}
J. Baschnagel, K. Binder and H.-P. Wittmann, J. Phys.: Condens. Matter
{\bf 5}, 1597 (1993).
%
\bibitem{lai90}
S. K. Lai and M. S. Lin, J. Non-Cryst. Solids {\bf 117/118}, 907
(1990).
%
\bibitem{miyagawa89}
H. Miyagawa and Y. Hiwatari, Phys. Rev. A {\bf 40}, 6007 (1989).
%
\bibitem{speedy94}
R. Speedy, Mol. Phys. {\bf 83}, 591 (1994).
%
\bibitem{kob95}
W. Kob, p. 1 in Vol. III of {\it Annual Reviews of Computational
Physics}, Ed. D. Stauffer (World Scientific, Singapore, 1995).
%
\bibitem{kob_lj}
W. Kob and H. C. Andersen, Phys. Rev. Lett. {\bf 73}, 1376 (1994);
%
Phys. Rev. E {\bf 51}, 4626 (1995);
%
{\it ibid.} {\bf 52}, 4134 (1995).
%
\bibitem{andersen80}
H. C. Andersen, J. Chem. Phys. {\bf 72}, 2384 (1980); 
%
see also Ref.~[3].
%
\bibitem{press92}
W. H. Press, S. A. Teukolsky, W. T. Vetterling and B. P. Flannery,
{\it Numerical Recipes}, (Cambridge University Press, Cambridge, 
1992).
%
\bibitem{vollmayr_phd}
K. Vollmayr, (PhD Thesis, Universit\"at Mainz, 1995).
%
\bibitem{vollmayr_sio2}
K. Vollmayr, W. Kob and K. Binder (preprint).
%
\bibitem{reinsch67}
C. H. Reinsch, Num. Math. {\bf 10}, 177 (1967);
C. H. Reinsch, {\it ibid.} {\bf 16}, 451 (1971).
%
\bibitem{angell81}
C. A. Angell, J. H. R. Clarke and L. V. Woodcock, Adv. Chem. Phys.
{\bf 48}, 397 (1981); see also C. A. Angell and L. M. Torell, J.
Chem. Phys. {\bf 78}, 937 (1983).
%
\bibitem{tool31}
A. Q. Tool and C. G. Eichlin, J. Opt. Soc. Amer. {\bf 14}, 276 (1931).
%
\bibitem{mct}
See, e.g., W. G\"otze and L.
Sj\"ogren, Rep. Prog. Phys. {\bf 55}, 241 (1992).
%
\bibitem{kob90}
W. Kob and R. Schilling, J. Phys. A: Math Gen. {\bf 23}, 4673 (1990)
and references therein.
%
\bibitem{grest_huse}
G. S. Grest, C. M. Soukoulis and K. Levin, Phys. Rev. Lett. {56}, 1148
(1986); 
%
G. S. Grest, C. M. Soukoulis, K. Levin and R. E. Randelman, p. 307 in {\em
Heidelberg Colloquium on Glassy Dynamics}, Vol 275 of {\em Lecture
Notes in Physics}, Ed. J.L. van Hemmen and I.  Morgenstern; Springer,
Berlin (1987);
%
D. A. Huse and D. S. Fisher, Phys. Rev. Lett. {\bf 57}, 2203 (1986).
%
\bibitem{new_optimization}
S. Cornell, K. Kaski and R. B. Stinchcombe, J. Phys. A: Math. Gen.
{\bf 24}, L865 (1991);
%
J. J\"ackle, R. B. Stinchcombe and S. Cornell, J. Stat. Phys. {\bf
62}, 425 (1991);
%
S. Shinomoto and Y. Kabashima, J. Phys. A: Math. Gen. {\bf 24}, L141
(1991);
%
P. Sibani, J. M. Pedersen K. H. Hoffmann and P. Salamon, Phys. Rev. A
{\bf 42}, 7080 (1990).
%
\bibitem{hansen_mcdo}
J.-P. Hansen and I. R. McDonald, {\it Theory of Simple Liquids}
(Academic, London, 1986).
%
\bibitem{vollmayr95}
K. Vollmayr, W. Kob and K. Binder, p. 117 in {\it Computer Simulation
Studies in Condensed Matter Physics VIII}, Ed. D. P. Landau, K. K. Mon
and H. B. Sch\"uttler, (Springer, Berlin, 1995).
%
\bibitem{clusters}
For a discussion on the nature of the nearest neighbor shells of
simple atomic liquids see J. Sietsma and B. J. Thiesse, Phys. Rev. B
{\bf 52}, 3248 (1995) and S. Cozzini and R. Ronchetti, (preprint
cond-mat/9511107).
%
\bibitem{vollmayr_epl}
K. Vollmayr, W. Kob and K. Binder, to be published in Europhysics
Lett.
%
\bibitem{physica}
See, e.g., in {\it Dynamics of Disordered Materials II}, Ed. A. J.
Dianoux, W. Petry and D. Richter, published in Physica A, {\bf 201}
(1993).
%
\bibitem{laird_schober}
B. B. Laird and H. R. Schober, Phys. Rev. Lett. {\bf 66}, 636 (1991);
H. R. Schober and B. B. Laird, Phys. Rev. B {\bf 44}, 6746 (1991).
%
\bibitem{olig_bermej}
H. R. Schober and C. Oligschleger, Nukleonika {\bf 39}, 185 (1994);
%
C. Oligschleger and H. R. Schober, Physica A {\bf 201}, 391 (1993);
%
C. Oligschleger, (PhD Thesis, J\"ulich, 1994);
%
H. R. Schober, C. Oligschleger and B. B. Laird, Journ. Non-Cryst.
Solids {\bf 156-158}, 965 (1993);
%
H. R. Schober and C. Oligschleger, (preprint Nov. 1995);
%
M. Garcia-Hernandez, F. J. Bermejo, B. F\aa k, J. L. Martinez, E.
Enciso, N. G. Almarza and A. Criado, Phys. Rev. B {\bf 48}, 149
(1993);
%
N. G. Almarza, E. Enciso and F. J. Bermejo, J. Chem. Phys. {\bf 99},
6876 (1993).
%
\bibitem{bembenek95}
S. D. Bembenek and B. B. Laird, Phys. Rev. Lett. {\bf 74}, 936
(1995).
%


\end{references}
\end{document}